# ODD-EVEN EMBEDDING SCHEME BASED MODIFIED REVERSIBLE WATERMARKING TECHNIQUE USING BLUEPRINT

Arijit Kumar Pal
B.Tech Student, Dept. of CSE
JIS College of Engineering
Kalyani, West Bengal, India
arijit1421@gmail.com

Poulami Das
M.Tech Scholar, Dept. of CSE
JIS College of Engineering
Kalyani, West Bengal, India
poulamidas1989@gmail.com

Nilanjan Dey
Asst. Prof. , Dept. of IT
JIS College of Engineering
dey.nilanjan@ymail.com

*Abstract* ------ **Digital watermarking is a technique of information adding or information hiding in order to identify the owner of the data in multimedia content. It seems that a signal or digital image can permanently embed over another digital data providing a good way to protect intellectual property from illegal replication. The cover data that is transmitted through the internet hides the watermark in a computer aided assertion method such that it becomes undetectable. Finally it stands as a hindrance over many operations without harming the embedded host document. Unfortunately, many owners of the digital materials such as images, text, audio and video are reluctant to the spreading of their documents on the web or other networked environment, because the ease of duplicating digital materials facilitates copyright violation. Digital media distribution occurs through various channels. The cover data may or may not hold any relation with the watermark information. In the last two decades, a considerable amount of research has been done on the digital watermarking of multimedia files such as audio, video, images and text. Different type of watermarking algorithms has been proposed by the researchers to achieve high level of security and authenticity.**
**In our proposed method, a modified reversible watermarking technique is introduced, which employs a blueprint generation of original image based on odd-even embedding methodology to yield large data hiding capacity, security as well as high watermarked quality. The experimental results demonstrate that, no matter how much secret data is embedded, the watermarked quality is about 51dB in this proposed scheme.**

## I. INTRODUCTION

The idea of hiding data in another media originated many years ago, as described in the case of steganography[1]. Nevertheless, technique of digital watermarking first appeared on 1993, when Tirkel et al. (1993) presented two techniques to hide data in images[2,3]. For building an effective water-marking algorithm the data must go through image-processing manipulations such as rotation, scaling, image compression and image enhancement. The term hiding refers to either making the messages undetectable or keeping their existence undisclosed. The contents within which information are embedded can be in digital formats, such as text, image, audio, video etc. Usually, image format is mostly used as the content for embedding due to its wide application. The image which is embedded within the content image is called watermark image and the content image is known as cover image. The combination of these two images is called watermarked image. Generally, the main concerns of data hiding techniques are the imperceptibility and hiding capacity. Imperceptibility is usually represented by PSNR (Peak Signal-to-Noise Ratio), which is used to measure the difference between the cover image and watermarked image.

When PSNR and correlation are considerably high, the cover image and the watermarked image cannot be distinguished by the naked eye.

Reversible watermarking technique is a special type of watermarking scheme useful for restoring the original information from the watermarked information. This feature is suitable for some important media such as medical and military media as these kinds of media does not allow any losses[4]. In this chapter we have included purposes of reversible watermarking, some techniques of reversible watermarking and progresses in this regard.

Generally a good watermarking scheme must meet the following requirements:
1. Robustness, 2. Imperceptibility, 3. Readily embedding and retrieving. Retrieved watermark by using reversible watermarking can be used to determine the ownership by comparing it with the assigned watermark. Similar to the conventional watermarking scheme, reversible watermarking also have to be robust against the intentional and unintentional attacks, imperceptible to avoid the attention of attacks and value lost.

In addition, this scheme of watermarking satisfies the following two requirements:
1. Blind (Reversible watermarking can retrieve the original information from the watermarked information directly without using the original data),
2. Higher Embedding Capacity (The required embedding capacity of the reversible watermarking schemes is much more than the conventional watermarking schemes as this scheme capable to embed the recovery-information and watermark information in to the original information).

There are some techniques of reversible watermarking scheme:
1. Even-odd method for reversible watermarking,
2. Alattar's Method for Reversible Watermarking[5,6],
3. Reversible Watermarking by Applying Data Comparison,
4. Reversible Watermarking using Difference Expansion[7],
5. Reversible Watermarking using Low Distortion Predication-error Expansion[8],
6. Reversible Watermarking by Histogram Bin Shifting,
7. Reversible Watermarking Based on Integer Wavelet Transformation,
8. Contrast Mapping.

All these schemes either have high hiding capacity and poor quality of watermarked image or good watermarked image quality and hiding capacity is low. In this paper, a novel reversible data hiding scheme is proposed.

The proposed scheme uses an Even-Odd embedding method to keep the watermarked image quality in a satisfied level and uses the multi-layer embedding technique to increase the hiding capacity. The experimental results demonstrate that the watermarked image quality is about 51dB in this proposed scheme no matter how much secret data is embedded.

The existed Even-Odd method embeds the message in the cover image according to the pixel value of cover image and message bit value[9]. Along with this location map is generated to restore the cover image. The generated location map tracks the pixel value change of the cover image.

## II. PROPOSED METHOD

The proposed modified Odd-Even embedding method embeds one-bit secret data into one pixel. The equation for embedding is different according to the pixel value which is even or odd and also to the message bit. And also a blueprint of the original image is built in the embedding phase. The blueprint is used to reconstruct the original pixel value after message extraction in extracting phase.

In the proposed scheme, the size (in pixel) of the message image has to be 1 lesser than the cover image. One bit secret data is embedded into every pixel of the cover image.

### A. Algorithm for Watermark Embedding

Input: Cover Image, Secret Data.
Output: Watermarked Image, Blueprint, Key Matrix.
Suppose that, C and S are the cover image and watermarked image and these images size is N x N respectively. B and K are the blueprint image and key image with size which size is N ×N. And M is the secret data, sized N ×⌊N-1⌋.
i and j are the coordinates that $0 \leq i \leq N - 1$, $0 \leq j \leq N-1$.
The steps of the embedding phase are as follows:

Step 1:
The first column of Watermarked image and Cover image is same.

---

```
1:    For i = 1 to N
2:        S (i, 1) = C (i, 1);
3:    End
```

---

Step 2:
The secret data is being embedded and the watermarked image is generated.

---

```
1:   For i = 1 to N
2:     For j = 1 to N-1
3:       If C (i, j+1) = odd and M (i, j) = 0
4:         S (i, j+1) = C (i, j+1);   //Direct Embedding
5:       End If
6:       If C (i, j+1) = even and M (i, j) = 1
7:         S (i, j+1) = C (i, j+1);   //Direct Embedding
8:       End If
9:       If C (i, j+1) = odd and M (i, j) = 1
10:         If C (i, j+1) = 255
11:           S (i, j+1) = C (i, j+1) - 1;   //Make it Even and Embed
12:         Else
13:           S (i, j+1) = C (i, j+1) + 1;
14:         End If
15:       End If
16:       If C (i, j+1) = even and M (i, j) = 0
17:         S (i, j+1) = C (i, j+1) + 1;   //Make it Odd and Embed
18:       End If
19:     End
20:   End
```

---

Step 3:
A blueprint of the cover image is also generated.

___

| | |
|---|---|
| 1: | *For i = 1 to N* |
| 2: | *For j = 1 to N* |
| 3: | *Temp = C (i, 1) – C (i, j);* |
| 4: | *B (i, j) = |Temp|;* |
| 5: | *If C (i, 1) <= C (i, j)* |
| 6: | *K (i, j) = 1;* |
| 7: | *Else* |
| 8: | *K (i, j) = 0;* |
| 9: | *End If* |
| 10: | *End* |
| 11: | *End* |

___

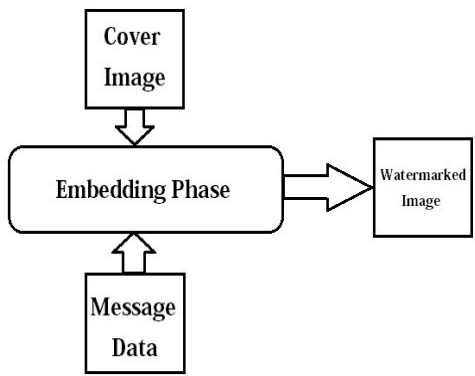     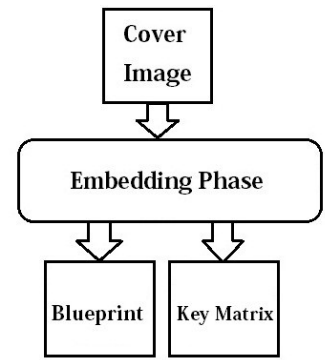

(a)                                                                                              (b)

**Fig1.** (a) Generating Watermark Image, (b) Generating Blueprint.

Let, the Cover Image(C) will be-

$$\begin{pmatrix} 225 & 225 & 227 & 228 \\ 226 & 226 & 228 & 229 \\ 226 & 224 & 225 & 226 \\ 221 & 224 & 228 & 228 \end{pmatrix}$$

and Message (M) –

$$\begin{pmatrix} 0 & 0 & 1 \\ 1 & 1 & 0 \\ 1 & 0 & 1 \\ 1 & 1 & 1 \end{pmatrix}_{(4 \times 3)}$$

[The message had embedded from the 2$^{nd}$ column of cover image, keeping the 1$^{st}$ column of cover image and watermarked image unchanged]

Considering the Position (1, 2) of cover image matrix-

C (1,2) = even and M (1,1) = 0

So, S (1,2) = C (1,2) + 1. Therefore, S (1,2) = 225 (Odd).

Thus, the Watermarked Image (S) will be–

$$\begin{pmatrix} 225 & 224 & 226 & 228 \\ 226 & 225 & 227 & 228 \\ 226 & 224 & 224 & 226 \\ 221 & 224 & 227 & 227 \end{pmatrix}_{(4 \times 4)}$$

Now, the Blue Print (B) will be –

$$\begin{pmatrix} 0 & 1 & 1 & 3 \\ 0 & 1 & 1 & 2 \\ 0 & 2 & 2 & 0 \\ 0 & 3 & 7 & 7 \end{pmatrix}$$

and the Key Matrix (K) will be-

$$\begin{pmatrix} 1 & 0 & 1 & 1 \\ 1 & 0 & 1 & 1 \\ 1 & 0 & 0 & 1 \\ 1 & 1 & 1 & 1 \end{pmatrix}$$

### B. Algorithm for Watermark Extraction –

Input: Watermarked Image, Blueprint, Key Matrix.
Output: Cover Image, Secret Data.

Suppose that, RC is the recovered cover image sized N x N and RM was the recovered secret data, sized N ×[N-1].

The steps of the extraction phase are as follows:

Step 1:

___________________________________________
*1:    For i = 1 to N*
*2:      For j = 1 to N-1*
*3:        If S (i, j+1) = odd*
*4:          RM (i, j) = 0;*
*5:        End If*
*6:        If S (i, j+1) = even*
*7:          RM (i, j) = 1;*
*8:        End If*
*9:      End*
*10:   End*
___________________________________________

Step 2:
       Reversibly the original cover image was made by the blueprint

___________________________________________
*1:      For i = 1 to N*
*2:        For j = 1 to N*
*3:          If K (i, j) = 1*
*4:            RC (i, j) = S (i, 1) + B (i, j);*
*5:          Else*
*6:            RC (i, j) = S (i, 1) - B (i, j);*
*7:          End If*
*8:        End*
*9:      End*
___________________________________________

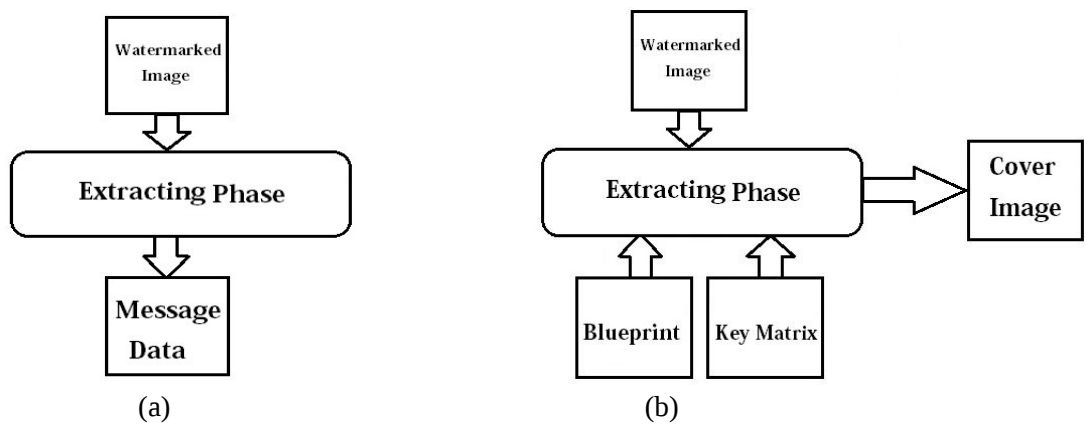

**Fig 2.**   (a) Message Recovering, (b) Cover Image Recovering.

Considering the Position (1, 2) of watermarked image –
S (1, 2) = Odd
So, RM (1, 1) = 0;   //As it's odd, a 0 was extracted.

Thus the Recovered Message (RM) Will Be-

$$\begin{pmatrix} 0 & 0 & 1 \\ 1 & 1 & 0 \\ 1 & 0 & 1 \\ 1 & 1 & 1 \end{pmatrix}_{(4 \times 3)}$$

Now, Considering the Position (1, 4) of watermarked image –

S (1, 4) = 228,   B (1, 4) = 3,   K (1, 4) = 1

  So, RC (1, 4) = S (1, 4) + B (1, 4) = 225+3 = 228

Thus, RC =

$$\begin{pmatrix} 225 & 224 & 226 & 228 \\ 226 & 225 & 227 & 228 \\ 226 & 224 & 224 & 226 \\ 221 & 224 & 227 & 227 \end{pmatrix}_{(4 \times 4)}$$

### III. RESULT AND DISCUSSIONS

The largest advantage of Even-Odd embedding method is the small degradation of the watermarked image after the secret data is embedded. The distortions that produced in the embedding phase can be balanced out in the next embedding phase, which is also the reason that twice Even-Odd embedding are used in one embedding layer. If the pixel value y is even /odd, the pixel value will become to y or y + 1 after the secret data is embedded. Therefore, no matter how many times Even-Odd embedding method is used, the maximum pixel value difference between the cover image and watermarked image will be 1. In the worst case, all the pixel values that used to embed are modified by 1 in the proposed scheme, then the PSNR is about 51dB, which means the quality of the watermarked image is always more than 51dB if Even-Odd embedding method is used. Due to the quality of the watermarked image is nearly changeless after more secret data is embedded, the multi-layer embedding method can be used to increase the hiding capacity.

**Table I.**

| Cover Image | Watermark | Watermarked Image | Extracted Watermark | PSNR between Cover Image and Watermarked Image | Correlation between Watermark and Extracted Watermark |
|---|---|---|---|---|---|
| 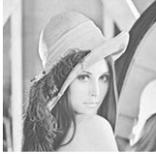 160x160 | 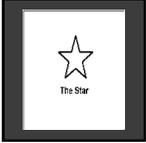 160x159 | 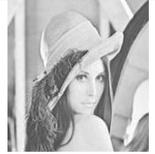 | 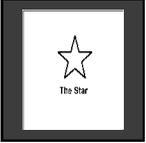 | 51.1435 | 1 |
| 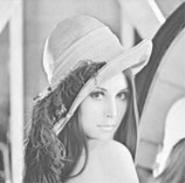 256x256 | 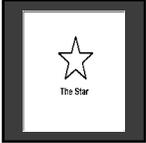 160x159 | 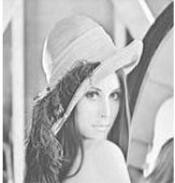 | 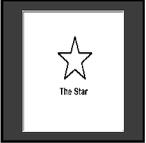 | 55.2127 | 1 |
| 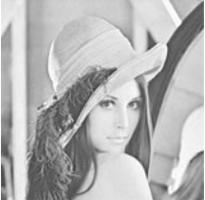 512x512 | 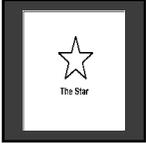 160x159 | 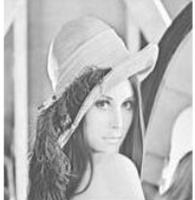 | 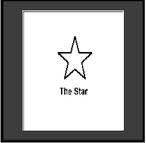 | 61.2482 | 1 |
| 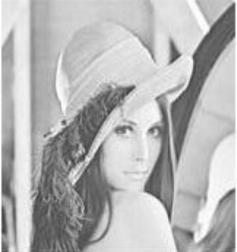 720x720 | 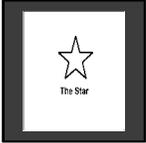 160x159 | 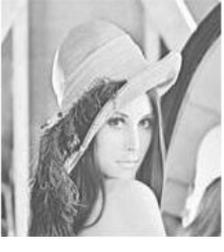 | 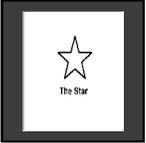 | 64.2111 | 1 |
| 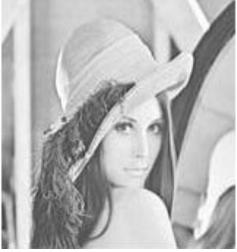 1080x1080 | 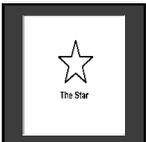 160x159 | 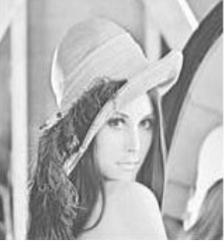 | 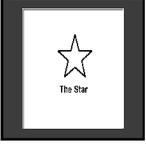 | 67.7214 | 1 |

*Peak Signal to Noise Ratio (PSNR)*

It measures the quality of a watermarked signal. This performance metric is used to determine perceptual transparency of the watermarked signal with respect to original signal[10]:

$$PSNR = \frac{XY \max_{x,y} P_{x,y}^2}{\sum_{x,y}(P_{x,y} - \overline{P}_{x,y})^2} \quad (1)$$

where, M and N are number of rows and columns in the input signal, $P_{x,y}$ is the original signal and $\overline{P}_{x,y}$ is the watermarked signal.

*Correlation Coefficient*

After secret image embedding process, the similarity of original signal x and watermarked signal x' is measured by the Standard Correlation Coefficient (c) as follows:

$$C = \frac{\sum_m \sum_n (x_{mn} - x')(y_{mn} - y')}{\sqrt{\left(\sum_m \sum_n (x_{mn} - x')^2\right)\left(\sum_m \sum_n (y_{mn} - y')^2\right)}} \quad (2)$$

where y and y' are the transforms of x and x'.

## IV. CONCLUSION

This proposed work is based on reversible watermarking scheme using Odd-Even embedding method. The possible distortions in the watermarked image of this proposed scheme were limited in a very low level as this embedding method could balance out the distortions that produced in the previous embedding phase. The experimental results demonstrated that the PSNR was about 51dB in the proposed scheme no matter how much secret data was embedded. In other words, the hiding capacity was infinite, which means the performance of the proposed scheme was excellent in both the PSNR and high capacity.

## REFERENCES

[1]. Ingemar, J.C., Matthew, L.M., Jeffrey, A.B., Jessica, F., Ton, K.: *Digital Watermarking and Steganography*, Morgan Kaufmann, Burlington,MA (2007)

[2]. Mintzer, *An Invisible Watermarking Technique for Image Verification* - Yeung, F – 1997.

[3]. O. Koval, S. Voloshynovskiy, T. Holotyak, and T. Pun, *Information-theoretic analysis of steganalysis in real images*, in *Proc. ACM Multimedia and Security Workshop 2006*, Geneva, Switzerland, 2006. [ps| bib].

[4]. DR .M A Dorairangaswamy, *A novel invisible and blind watermarking scheme for copyright protection of digital images*, International journal of computer science and network security vol9 No.4 ,April 2009.

[5]. Nilanjan Dey, Poulami Das, Achintya Das, Sheli Sinha Chaudhuri, *Feature Analysis for the Blind–Watermarked Electroencephalogram Signal in Wireless Telemonitoring Using Alattar's Method,* 5th International Conference on Security of Information and Networks (SIN 2012), In Technical Cooperation with ACM Special Interest Group on Security, Audit and Control (SIGSAC). Proceedings by ACM Press and Digital Library, 22-26 October 2012, Jaipur, India**.**